\documentclass[12pt]{article}
\usepackage{a4wide,epsfig,amsmath,amssymb,cite,scalefnt,graphicx}

\def\npb{{\it{Nucl.\ Phys.}\ }{\bf B}}
\def\plb{{{\it Phys.\ Lett.}\ }{ \bf B}}
\def\be{\begin{equation}}
\def\ee{\end{equation}}
\def\bea{\begin{eqnarray}}
\def\eea{\end{eqnarray}}
\def\nn{\nonumber\\}

\def\tr{\hbox{tr}}

\def\prd{{{\it Phys.\ Rev.}\ }{\bf D}}

\def\thetabar{{\overline\theta}}

\def\psibar{\overline{\psi}}

\def\Tr{\mathop{\rm Tr}}

\def\lambdabar{\overline\lambda}

\def\Dbar{\overline D}

\def\psibar{\overline\psi}

\def\Ytil{\tilde Y}

\def\Ncal{{\cal N}}

\def\Ytil{\tilde Y}

\def\pa{\partial}
\def\pabar{\overline\partial}

\input epsf
\begin{document}

\begin{titlepage}
\begin{flushright}
LTH901\\
\end{flushright}
\date{}
\vspace*{3mm}

\begin{center}
{\Huge
Exact $\beta$-functions in softly-broken $\Ncal=2$ Chern-Simons matter
theories}\\[12mm]
{\bf I.~Jack and C.~Luckhurst}\\
%\end{center}

\vspace{5mm}
Dept. of Mathematical Sciences,
University of Liverpool, Liverpool L69 3BX, UK\\

\end{center}

\vspace{3mm}
\begin{abstract}
We present exact results for the $\beta$-functions for
the soft-breaking parameters in softly-broken $\Ncal=2$ Chern-Simons matter
theories in terms of the anomalous dimension in the unbroken theory. We 
check our results explicitly up to the two loop level.

\end{abstract}

\vfill

\end{titlepage}

Chern-Simons gauge theories have attracted attention for a considerable 
time due to their topological nature\cite{schwarza, witten, djt}
(in the pure gauge case) and their 
possible relation to the quantum Hall effect and high-$T_c$ 
superconductivity. More recently  
there has been substantial interest in $\Ncal=2$ supersymmetric Chern-Simons
matter theories in the context of the AdS/CFT 
correspondence (see Refs.~\cite{penatia,ASW,penatib} for 
details and a comprehensive list of references), and it therefore seems 
timely to
consider the softly-broken version of the theory. It is already well-known
that the $\beta$-functions for the soft-breaking parameters in 
softly-broken $\Ncal=1$ supersymmetric gauge theories in four dimensions may be
expressed exactly in terms of the anomalous dimensions and gauge 
$\beta$-function for the unbroken theory. (See Ref.~\cite{jjw} for a
complete description of the most general case.) Moreover this 
leads\cite{jjrginv} to exact 
renormalisation group invariant solutions for the soft-breaking parameters--the
``anomaly-mediated supersymmetry-breaking'' (AMSB) 
solutions\cite{lr}-\cite{kss}. 
The purpose of this note is to point
out that similar results hold for $\Ncal=2$ Chern-Simons matter theories  in 
three dimensions; indeed the results are simpler due to the absence of a  
gauge coupling (which reflects the topological nature of the gauge part of the
theory).

Our results are based on a set of rules devised by Yamada\cite{yam}
for obtaining the $\beta$-functions
for the scalar soft-breaking couplings (in four dimensions) starting from the 
anomalous 
dimension for the chiral superfields. We shall present here an abridged
derivation based on Ref.~\cite{jjgaug}; see Ref.~\cite{jjw} for the complete
version. Yamada's rules are based on 
the spurion formalism\cite{spur}, which enables one to write 
the softly broken $N=2$ theory in terms of superfields. 
The lagrangian for the theory can be written
\be
L=L_{SUSY}+L_{SB}+L_{GF}+L_{FP} 
\label{Aa}
\ee
where $L_{SUSY}$ is the usual $N=2$ supersymmetric lagrangian\cite{ivanov}, 
\bea
L_{SUSY}=&\int d^3x\int d^4\theta\left(
2k\int_0^1dt\Tr[\Dbar^{\alpha}(
e^{-tV}D_{\alpha}e^{tV})]+\Phi^j (e^{V_AR_A})^i{}_j\Phi_i\right)\nn
&+\left(\int d^3x\int d^2\theta W(\Phi)
+\hbox{h.c.}\right),
\eea
where $V$ is the vector
superfield, $\Phi$ the chiral matter superfield and where the 
superpotential $W(\Phi)$ is given by
\be
W(\Phi)=\frac{1}{4!}Y^{ijkl}\Phi_i\Phi_j\Phi_k\Phi_l
+\frac{1}{3!}Z^{ijk}\Phi_i\Phi_j\Phi_k
+\frac{1}{2!}\mu^{ij}\Phi_i\Phi_j.
\ee
(We use the 
convention that $\Phi^i=(\Phi_i)^*$.) 
We assume a simple gauge group; a gauge group with a $U(1)$ factor could
also include a linear term in the superpotential.
Gauge invariance requires the gauge coupling $k$ to be 
quantised, so that $2\pi k$ is an integer. The vector superfield $V$ is in the
adjoint representation, $V=V_AT_A$ where $T_A$ are the generators of the
fundamental representation, satisfying 
\bea
[T_A,T_B]=&if_{ABC}T_C,\\
\Tr(T_AT_B)=&\frac12\delta_{AB}.
\eea
The chiral superfield can be in a general representation, with gauge
matrices denoted $R_A$ satisfying
\bea
[R_A,R_B]=&if_{ABC}R_C,\\
\Tr(R_AR_B)=&T(R)\delta_{AB}.
\label{Tdef}
\eea
In three dimensions the Yukawa couplings $Y^{ijkl}$ are dimensionless 
and the theory is renormalisable. 
The soft breaking part $L_{SB}$ may be written\cite{GN}
\bea
L_{SB}&=\int d^2\theta\eta\left(\frac{1}{4!}h^{ijkl}
\Phi_i\Phi_j\Phi_k\Phi_l+\frac{1}{3!}g^{ijk}\Phi_i\Phi_j\Phi_k
+\frac{1}{2!}b^{ij}\Phi_i\Phi_j+\hbox{h.c.}\right)\nn
&\quad -\int d^4\theta\eta^*\eta\Phi^j(m^2)^i{}_j(e^{V_AR_A})_i{}^k\Phi_k,
\label{Aba}
\eea
where $\eta=\theta^2$ is the spurion external field. Note that in three
dimensions there is no soft term corresponding to the 
four-dimensional gaugino mass term.
The gauge-fixing and Fadeev-Popov terms are contained in $L_{GF}$ and
$L_{FP}$ respectively.  
It is convenient to introduce a generalised form $\gamma_{\eta}$ 
of the anomalous 
dimension $\gamma$ of the chiral supermultiplet, given by: 
\be
\gamma_{\eta}=\gamma+\gamma_1\eta+\gamma_1^{\dagger}\eta^*+\gamma_2
\eta^*\eta.
\label{Ah}
\ee
It was shown by Yamada\cite{yam} that $(\gamma_{\eta})^i{}_j$ could be obtained
from $(\gamma)^i{}_j$ by the following rules (simpler in three than in four
dimensions due to the absence of a running gauge coupling):
\begin{enumerate}
\item Replace $Y^{lmno}$ by $Y^{lmno}-h^{lmno}\eta$.
\item Insert $\delta^{l'}{}_l+(m^2)^{l'}{}_l\eta^*\eta$ between contracted 
indices $l$ and $l'$ in $Y$ and $Y^*$, respectively: $Y^{lmno}Y_{lm'n'o'}
\rightarrow Y^{lmno}Y_{lm'n'o'}+Y^{lmno}(m^2)^{l'}{}_lY_{l'm'n'o'}
\eta^*\eta
$ (where,
here and subsequently, $Y_{lmno}=(Y^{lmno})^*$).
\item Replace a term $T^i{}_j$ in 
$\gamma^i{}_j$ with no Yukawa couplings by 
$T^i{}_j-(m^2)^i{}_kT^k{}_j\eta^*\eta$.
\end{enumerate}
\noindent$\gamma_1$ and $\gamma_2$ may then 
be obtained by extracting the coefficients of $\eta$
and $\eta^*\eta$ respectively.
In the case of $\gamma_1$,  
the above rules can be subsumed by the simple relation 
\be
(\gamma_1)^i{}_j={\cal O}\gamma^i{}_j,
\label{Aj}
\ee
where
\be
{\cal O}=-h^{lmno}{\partial
\over{\partial Y^{lmno}}}.
\label{Ajb}
\ee
It is  straightforward to show that 
\be
\beta_h^{ijkl}=\gamma^{(i}{}_mh^{jkl)m}-2\gamma_1^{(i}{}_mY^{jkl)m}.
\label{Ai}
\ee
This result is similar in form to the standard result for $\beta_Y$
which follows from the non-renormalisation theorem (which is valid 
for $\Ncal=2$ supersymmetric theories in three dimensions), namely
\be
\beta_Y^{ijkl}=\gamma^{(i}{}_mY^{jkl)m}.
\label{Aia}
\ee
We also have the analogous results for the soft couplings corrresponding
to dimensionful supersymmetric couplings,
\bea
\beta_g^{ijk}=&\gamma^{(i}{}_mg^{jk)m}-2\gamma_1^{(i}{}_mZ^{jk)m},\\
\beta_b^{ij}=&\gamma^{(i}{}_mb^{j)m}-2\gamma_1^{(i}{}_m\mu^{j)m}.
\label{Aid}
\eea
It also follows from Eqs.~(\ref{Aba}) and (\ref{Ah}) that
\be
(\beta_{m^2})^i{}_j=\frac12\gamma^i{}_k(m^2)^k{}_j
+\frac12(m^2)^i{}_k\gamma^k{}_j+\gamma_2^i{}_j,
\label{Aib}
\ee
which we may write using Yamada's rules as 
\be
(\beta_{m^2})^i{}_j=\left[2{\cal O}{\cal O}^*
+\Ytil_{lmn}{\partial\over{\partial Y_{lmn}}}
+\Ytil^{lmn}{\partial\over{\partial Y^{lmn}}}\right]\gamma^i{}_j,
\label{Ajc}
\ee
where
\be
\Ytil^{ijkl}=(m^2)^i{}_mY^{mjkl}+(m^2)^j{}_mY^{imkl}+(m^2)^k{}_mY^{ijml}
+(m^2)^l{}_mY^{ijkm}.
\label{Ajd}
\ee
The exact results Eqs.~(\ref{Ai}) and (\ref{Ajc}) for the $\beta$-functions
lead to exact renormalisation group invariant solutions for the soft-breaking
couplings, namely 
\bea
h^{ijkl}=-M_0\beta_Y^{ijkl},\\
g^{ijk}=-M_0\beta_Z^{ijk}+\kappa_1Z^{ijk},\\
b^{ij}=-M_0\beta_{\mu}^{ij}+\kappa_2\mu^{ij},\\
(m^2)^i{}_j=\frac12|M_0|^2\mu\frac{d\gamma^i{}_j}{d\mu},
\label{traj}
\eea   
where $M_0$, $\kappa_1$, $\kappa_2$ are constant masses.
These results can be proved following the four-dimensional discussion in
Ref.~\cite{jjrginv} (though the terms with $\kappa_{1,2}$ were given for
the first time in Ref.~\cite{jjh});
but once more the details are simpler due to the 
non-running of the gauge coupling. We note that in the case 
of a gauge group with a $U(1)$ factor and a linear term in the superpotential,
additional terms are expected\cite{jjw} in the expressions for $\beta_g$ and
$\beta_b$ in Eq.~(\ref{Aid}), and thence corresponding extra terms
in Eqs.~(\ref{traj}); there should also be an exact expression
for the $\beta$-function corresponding to the linear soft coupling, and
an exact RG-invariant solution for this coupling. There is also 
potentially an additional term\cite{jjxi} 
in the solution for $m^2$ corresponding
to the possible Fayet-Iliopoulos term.

We now turn to our check of the results Eqs.~(\ref{Ai}) and (\ref{Ajc}) up
to two loops using the component formulation of the theory (there are 
no divergences at odd loop orders for a theory in odd dimensions, so this 
is the simplest non-trivial check).
The first ingredient is 
the anomalous dimension of the chiral superfield, which is given at two loops
by
\be
64\pi^2\gamma^{(2)}=
\frac13Y_2-2k^{-2}C_2(R)C_2(R)-k^{-2}T(R)C_2(R)+k^{-2}C_2(G)C_2(R)
\label{gamphi}
\ee  
where
\bea
(Y_2)^i{}_j=&Y^{iklm}Y_{jklm}\\
C_2(R)=&R_AR_A,\\
C_2(G)\delta_{AB}=&f_{ACD}f_{BCD}\\
\eea
and $T(R)$ is defined in Eq.~(\ref{Tdef}).
This result may readily be obtained by $\Ncal=2$ superfield 
methods\cite{agk,GN,ASW,penatib}; see
the appendix for the $\Ncal=2$ superfield conventions.

An expression for the two-loop anomalous dimension for an $\Ncal=1$ 
theory in three dimensions (with no Yukawa coupling) 
is given in Ref.~\cite{akk}. This does not agree
with the $k^{-2}$ terms in Eq.~(\ref{gamphi}) when specialised 
to the $\Ncal=2$ case. Presumably 
this is because the result is in general gauge-dependent and the 
$\Ncal=1$ and $\Ncal=2$ Feynman gauges are not equivalent. Since $\Ncal=2$
supersymmetry is not manifest in the $\Ncal=1$ formalism, one would not
expect Eq.~(\ref{Aia}) to be valid using the anomalous dimension computed using
the $\Ncal=1$ formalism. We have however checked explicitly via a component
calculation that the $\beta$ function for the Yukawa coupling is indeed given
by Eq.~(\ref{Aia}) with the anomalous dimension of Eq.~(\ref{gamphi}). 

We then find from Eq.~(\ref{Aj}) that 
\be
64\pi^2(\gamma^{(2)}_1)^i{}_j=-\frac13h^{ilmn}Y_{jlmn}
\ee
and that therefore (using Eq.~(\ref{Ai}))
\bea
64\pi^2\beta_h^{ijkl(2)}=&[\frac13Y_2
-2k^{-2}C_2(R)C_2(R)-k^{-2}T(R)C_2(R)+k^{-2}C_2(G)C_2(R)]^i{}_m
h^{mjkl}\nn
&+\frac23h^{ilmn}Y_{plmn}Y^{pjkl}+\hbox{cyclic perms}.
\label{htwo}
\eea
We also find from Eq.~(\ref{Ajc}) that 
\be
64\pi^2(\beta_{m^2})^i{}_j=\frac23h^{iklm}h_{jklm}
+\frac13(m^2)^i{}_k(Y_2)^k{}_j+\frac13(Y_2)^i{}_k(m^2)^k{}_j\nn
+2Y^{iklm}(m^2)^{k'}{}_kY_{jk'lm}.      
\label{mtwo}
\ee

It is straightforward to verify these results by a component calculation.
The supersymmetric Lagrangian is given in components by\cite{schwarz}
\bea
L_{SUSY}=&L_{CS}+L_{m}\\
L_{CS}=&2k\Tr[\epsilon^{\mu\nu\rho}(A_{\mu}\pa_{\nu}A_{\rho}+\frac{2i}{3}
A_{\mu}A_{\nu}A_{\rho})-\lambdabar\lambda+2D\sigma]\\
L_m=&D_{\mu}\phi^i D^{\mu}\phi_i+i\psibar^i\gamma^{\mu}D_{\mu}\psi_i
+F^iF_i\nn
-&\phi^i\sigma^2\phi_i+\phi^iD\phi_i+i\phi^{\dagger}\lambdabar\psi
-i\psibar\lambda\phi\nn
&+\left(\frac{1}{3!}Y^{ijkl}\phi_i\phi_j\phi_kF_l
+\frac14Y^{ijkl}\phi_i\phi_j\psibar_k\psi_l+\hbox{h.c.}\right),
\eea
where $\lambda$ and $\psi$ are two-component Dirac spinors,
$\lambdabar=\lambda^{\dagger}\gamma_0$, $D_{\mu}=\pa_{\mu}+iA_{\mu}$ and 
we have set $\mu^{ij}=Z^{ijk}=0$ for simplicity, in order to focus on the 
dimensionless couplings.
After eliminating the auxiliary fields $D$, $\sigma$ we obtain
\bea
L_{CS}=&\Tr[\epsilon^{\mu\nu\rho}(A_{\mu}\pa_{\nu}A_{\rho}+\frac{2i}{3}
A_{\mu}A_{\nu}A_{\rho})],\\
L_m=&D_{\mu}\phi^i D^{\mu}\phi_i+i\psibar^i\gamma^{\mu}D_{\mu}\psi_i
\nn
-&(\phi^{\dagger}R_A\phi)(\phi^{*}R_B\phi)(\phi^{*}R_AR_B\phi)
+(\phi^{*}R_A\phi)(\psibar^{*}R_A\psi)
+2(\psibar^{*}R_A\phi)(\phi^{*}R_A\psi)\nn
-&\frac{1}{(3!)^2}Y^{ijkn}Y_{i'j'k'n}\phi_i\phi_j\phi_k\phi^{i'}
\phi^{j'}\phi^{k'}+(\frac14Y^{ijkl}\phi_i\phi_j\psibar_k\psi_l+\hbox{h.c.})
\eea
The soft-breaking lagrangian is given by
\be
L_{SB}=-\left(\frac{1}{4!}h^{ijkl}\phi_i\phi_j\phi_k\phi_l+\hbox{h.c.}\right)
-(m^2)^i{}_j\phi_i\phi^j,
\ee
where we have set $b^{ij}=g^{ijk}=0$.

The diagrams contributing to the anomalous dimension of the scalar component 
field $\phi$ at two loops are depicted in Fig.~1, with
scalar, fermion, gauge and ghost propagators denoted by dashed, unbroken,
wavy and dotted lines respectively. We work in a standard Feynman gauge in 
components (see the Appendix). The divergent 
contributions from the diagrams in Fig. 1 to 
$\pa_\mu\phi^{*}\pa^{\mu}\phi$ are given by (using dimensional
regularisation and working in $d=3-\epsilon$ dimensions)
\bea
L\Gamma^{(2)}_{\phi(a)}=&\frac13Y_2+\frac16k^{-2}[4C_2(R)-2C_2(G)+5T(R)]C_2(R),\\
L\Gamma^{(2)}_{\phi(b)}=&\frac{1}{12}k^{-2}[-4C_2(R)+C_2(G)]C_2(R),\\
L\Gamma^{(2)}_{\phi(c)}=&-\frac23k^{-2}T(R)C_2(R),\\
L\Gamma^{(2)}_{\phi(d)}=&-\frac23k^{-2}T(R)C_2(R),\\
L\Gamma^{(2)}_{\phi(e)}=&\frac16k^{-2}C_2(G)C_2(R),\\
L\Gamma^{(2)}_{\phi(f)}=&-\frac16k^{-2}C_2(G)C_2(R),\\
L\Gamma^{(2)}_{\phi(g)}=&\frac23k^{-2}[-2C_2(R)+C_2(G)]C_2(R),\\
L\Gamma^{(2)}_{\phi(h)}=&\frac13k^{-2}C_2(G)C_2(R),
\eea
where $L=64\pi^2\epsilon$, leading to 
\be
\gamma^{(2)}_{\phi}=\frac13Y_2-k^{-2}C_2(R)C_2(R)-\frac12k^{-2}T(R)C_2(R)
+\frac34k^{-2}C_2(G)C_2(R)
\label{gamphia}
\ee  
which agrees (up to an overall factor of 4, whose origin we have not been
able to identify) 
with the component-field calculation in Ref.~\cite{akk}, when the 
relevant result is specialised
to the case of $\Ncal=2$ supersymmetry. Note that since there are no
simple poles at one loop, there are no double poles at two loops and no need
to consider diagrams with counterterm insertions at this order.
The list of diagrams contributing to $\beta_h$ and $\beta_{m^2}$ can be 
shortened by noting that any logarithmically divergent diagram where an 
external scalar emerges from a $\phi^{*}A\phi$ vertex is zero by symmetry, due
to the form of the gauge propagator (see the Appendix).
The diagrams contributing to the two-loop $\beta$ functions for $m^2$ and 
$h$ are shown in Figs.~2 and 3 respectively. They yield divergent 
contributions to the effective action given by
\bea
L\Gamma_{m^2(a)}^{(2)}
=&\{Y^{iklm}(m^2)^{k'}{}_kY_{jk'lm}
+\frac12k^{-2}[4C_2(R)-2C_2(G)+T(R)]C_2(R)(m^2)^i{}_j\}\phi_i\phi^j\nn
&+2k^{-2}\tr[R_AR_Bm^2]\phi^{*}R_AR_B\phi,\\
L\Gamma_{m^2(b)}^{(2)}=&-\frac14k^{-2}[4C_2(R)-C_2(G)]C_2(R)\phi^{*}m^2\phi,\\
L\Gamma_{m^2(c)}^{(2)}=&-2k^{-2}\tr[R_AR_Bm^2]\phi^{*}R_AR_B\phi,\\
L\Gamma_{m^2(d)}^{(2)}=&\frac13h^{iklm}h_{jklm}\phi_i\phi^j,
\eea
and
\bea
L\Gamma_{h(a)}^{(2)}=&\frac14k^{-2}[h^{ijmn}(R_AR_B)^k{}_m(R_AR_B)^l{}_n
-\frac{1}{12}h^{ijkm}[C_2(G)C_2(R)]^l{}_m]\phi_i\phi_j\phi_k\phi_l,\\
L\Gamma_{h(b)}^{(2)}=&\frac14k^{-2}[
-2h^{ijmn}(R_AR_B)^k{}_m(R_AR_B)^l{}_n\nn
&+\frac16h^{ijkm}\{4C_2(R)C_2(R)+T(R)C_2(R)\}^l{}_m]\phi_i\phi_j\phi_k\phi_l,\\
L\Gamma_{h(c)}^{(2)}=&\{\frac13h^{ilmn}Y_{plmn}Y^{pjkl}
+\frac14k^{-2}h^{ijmn}(R_AR_B)^k{}_m(R_AR_B)^l{}_n\nn
&-\frac{1}{12}k^{-2}h^{ijkm}[C_2(R)C_2(R)]^l{}_m\}\phi_i\phi_j\phi_k\phi_l.
\eea
These add to 
\bea
L\Gamma_{m^2}^{(2)}=&
\{Y^{iklm}(m^2)^{k'}{}_kY_{jk'lm}+\frac13h^{iklm}h_{jklm}\phi_i\phi^j\nn
&+\frac14k^{-2}[4C_2(R)-3C_2(G)+2T(R)]C_2(R)(m^2)^i{}_j\}\phi_i\phi^j
\label{mtwoa}
\eea
and 
\bea
L\Gamma_h^{(2)}=&\frac16\{\frac13h^{ilmn}Y_{plmn}Y^{pjkl}\nn
&-\frac{1}{8}k^{-2}h^{ijkm}[2T(R)C_2(R)+4C_2(R)C_2(R)
-C_2(G)C_2(R)]^l{}_m\}\phi_i\phi_j\phi_k\phi_l.
\label{htwoa}
\eea
We expect from elementary renormalisation theory 
that the soft-breaking $\beta$-functions will satisfy
\bea
2L\Gamma_h^{(2)}=&\frac{1}{4!}
\left(\beta_h^{ijkl(2)}-4(\gamma^{(2)}_{\phi})^l{}_mh^{ijkm}\right)
\phi_i\phi_j\phi_k\phi_l,\nn
2L\Gamma_{m^2}^{(2)}=&(\beta^{(2)}_{m^2})^i{}_j\phi^j\phi_i
-(\gamma^{(2)}_{\phi}m^2)^i{}_j\phi^j\phi_i
-(m^2\gamma^{(2)}_{\phi})^i{}_j\phi^j\phi_i,
\eea
writing the results in this form to avoid cumbersome symmetrisations.
We easily verify these identities using Eqs.~(\ref{mtwo}), (\ref{htwo}),
(\ref{mtwoa}), (\ref{htwoa}), (\ref{gamphia}).

The maximal supersymmetry for a Chern-Simons theory with a single gauge 
group is $\Ncal=3$. The component formulation of this theory was presented
in Ref.~\cite{kao}. The quantum properties of this theory were discussed in
Ref.~\cite{buch} based on the $d=3$ $\Ncal=3$ harmonic superspace 
formalism developed in Ref.~\cite{zupnik}, and it was shown that this theory 
is all-orders finite. It would be interesting to investigate whether the 
softly-broken version of this theory is also finite. 
Theories with higher degrees of 
supersymmetry (up to $\Ncal=8$)\cite{bagger} 
may be obtained in the case of direct product groups and 
matter in the bi-fundamental representation. A rich variety of these
theories\cite{bkks}-\cite{HLT} are 
expected to be superconformal by virtue of the
$\hbox{AdS}_4$/$\hbox{CFT}_3$ correspondence, originally
stated in Ref~\cite{ABMJ}. These theories can be
expressed in terms of $\Ncal=2$ superfields and are obtained by a  
judicious choice of field content and also a particular choice of 
Yukawa couplings (as a function of the gauge couplings). 
The conformal properties of a range of these models was checked explicitly
at the two-loop level in Refs.~\cite{ASW,penatib}. It would be quite 
straightforward to extend our results to the case of direct product gauge groups
and thereby derive exact results for the softly broken versions   
of these theories. One could then ask whether there were a choice of soft 
couplings which would maintain finiteness. In the case of $\beta_h$ this
would entail arranging for $\gamma_1$ to vanish; this is not guaranteed
by the vanishing of $\gamma$, since 
the derivative in Eq.~(\ref{Ajb}) would be taken before specialising to
the special form for the Yukawa couplings which guarantees the extended
supersymmetry. Nevertheless it was shown in the four-dimensional 
case\cite{jjrginv} that there was a choice of soft couplings
which would guarantee $\gamma_1=0$. However this relied on the existence 
of the gaugino mass as a soft coupling and a similar choice is not possible 
here; there is therefore no obvious way to guarantee the vanishing 
of $\beta_h$. The same 
argument applied to Eq.~(\ref{Aib}) would imply that we could not
render $\beta_{m^2}$ zero. The softly-broken versions of these 
superconformal theories would therefore not be finite.

Finally, it would be interesting to address the question of gauge groups with 
a $U(1)$ factor, where, as we have noted, there are additional technical
subtleties.  

\bigskip

\bigskip

\noindent
{\large{\bf Acknowledgements\\}}
One of us (CL) was supported by a University of Liverpool studentship. IJ is 
grateful for useful discussions with Tim Jones.

\bigskip

\noindent
{\large{\bf{Appendix}\hfil}}

In this appendix we list our superspace and supersymmetry conventions.
We use a metric signature $(+--)$ so that a possible choice of $\gamma$ matrices
is $\gamma^0=\sigma_2$, $\gamma^1=i\sigma_3$, $\gamma^2=i\sigma_1$
with
\be
(\gamma^{\mu})_{\alpha}{}^{\beta}=(\sigma_2)_{\alpha}{}^{\beta}, 
\ee
etc. We then have 
\be
\gamma^{\mu}\gamma^{\nu}=\eta^{\mu\nu}-i\epsilon^{\mu\nu\rho}\gamma_{\rho}.
\ee
We have\cite{penatib}
 two complex two-spinors $\theta^{\alpha}$ and $\theta^{\alpha}$ 
with indices raised and lowered according to
\be
\theta^{\alpha}=C^{\alpha\beta}\theta_{\beta},\quad
\theta_{\alpha}=\theta^{\beta}C_{\beta\alpha},
\ee
with $C^{12}=-C_{12}=i$. We then have
\be
\theta_{\alpha}\theta_{\beta}=C_{\beta\alpha}\theta^2,\quad
\theta^{\alpha}\theta^{\beta}=C^{\beta\alpha}\theta^2,
\ee
where 
\be
\theta^2=\frac12\theta^{\alpha}\theta_{\alpha}.
\ee
The supercovariant derivatives are defined by
\bea
D_{\alpha}=&\pa_{\alpha}+\frac{i}{2}\theta^{*\beta}\pa_{\alpha\beta},\\
\Dbar_{\alpha}=&\pabar_{\alpha}+\frac{i}{2}\theta^{\beta}\pa_{\alpha\beta},
\eea
where
\be
\pa_{\alpha\beta}=\pa_{\mu}(\gamma^{\mu})_{\alpha\beta},
\ee
satisfying
\be
\{D_{\alpha},\Dbar_{\beta}\}=i\pa_{\alpha\beta}.
\ee
(We have used the notation $\theta^*$ rather than the usual $\thetabar$ to avoid
confusion with $\lambdabar$ defined earlier in the component formulation.) 
We also define
\be
d^2\theta=\frac12d\theta^{\alpha}d\theta_{\alpha}
\quad d^2\theta^*=\frac12d\theta^{*\alpha}d\theta^*_{\alpha},
d^4\theta=d^2\theta d^2\theta^*,
\ee
so that
\be
\int d^2\theta\theta^2=\int d^2\theta^*\theta^{*2}=-1.
\ee
The vector superfield $V(x,\theta,\theta^*)$ is expanded in Wess-Zumino gauge
as
\be
V=i\theta^{\alpha}\theta^*_{\alpha}\sigma+\theta^{\alpha}\theta^{*\beta}
A_{\alpha\beta}-\theta^2\theta^{*\alpha}\lambda^*_{\alpha}
-\theta^{*2}\theta^{\alpha}\lambda_{\alpha}+\theta^2\theta^{*2}D,
\ee
and the chiral field is expanded as
\be
\Phi=\phi(y)+\theta^{\alpha}\psi_{\alpha}(y)-\theta^2F(y),
\ee
where
\be
y^{\mu}=x^{\mu}+i\theta\gamma^{\mu}\theta^*.
\ee
The scalar, fermion and gauge propagators $\Delta_S$, $\Delta_F$
and $\Delta_V$ are given by (using a standard Feynman-type gauge) 
\be 
\Delta_S=\frac{1}{k^2},\quad \Delta_F=\frac{k_{\mu}\gamma^{\mu}}{k^2},
\quad (\Delta_V)^{\mu\nu}=\frac{i\epsilon^{\mu\nu\rho}k_{\rho}}{k^2}.
\ee

\begin{figure}
\includegraphics{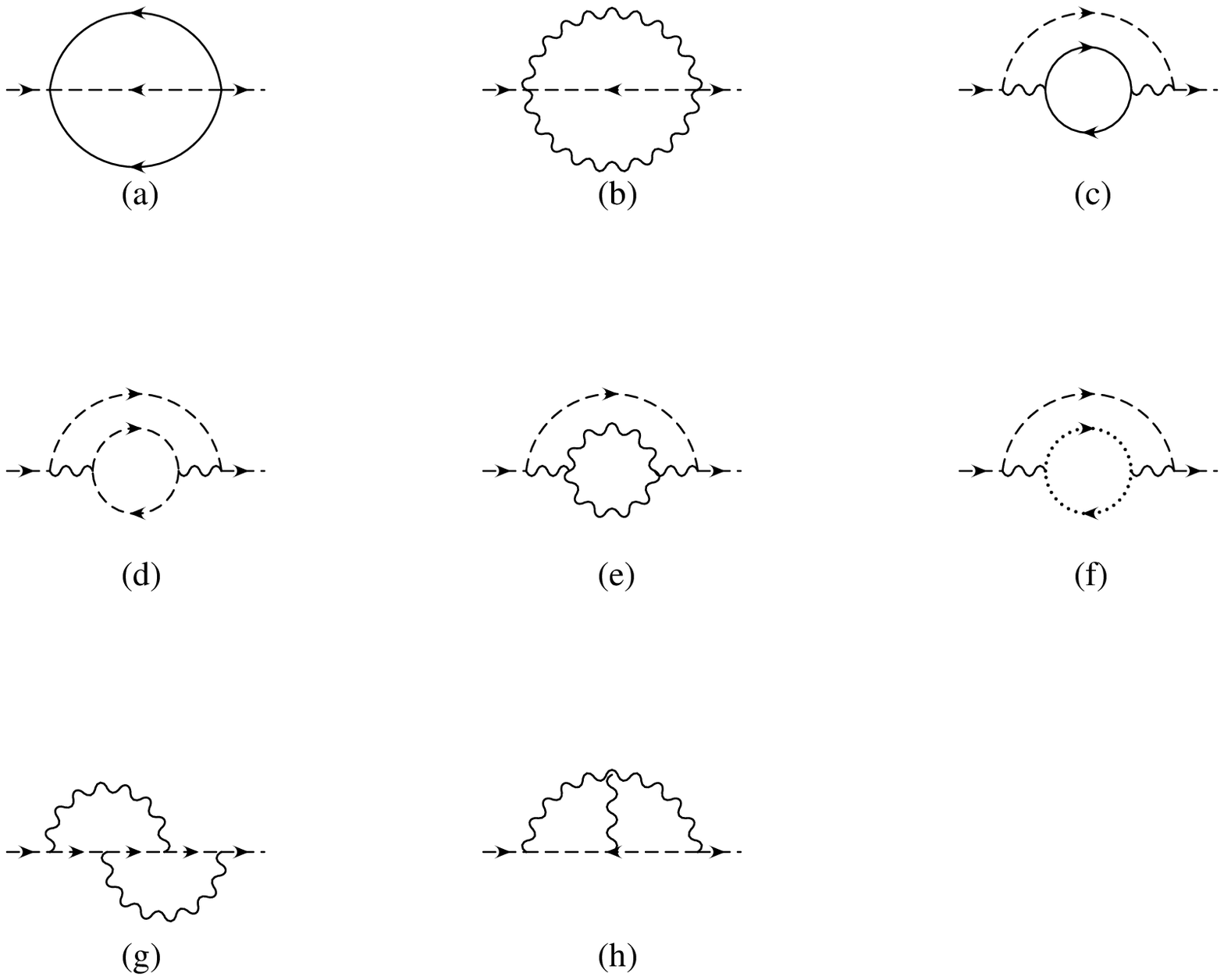}
\caption{Diagrams contributing to $\gamma_{\phi}^{(2)}$}
\end{figure}

\begin{figure}
\includegraphics{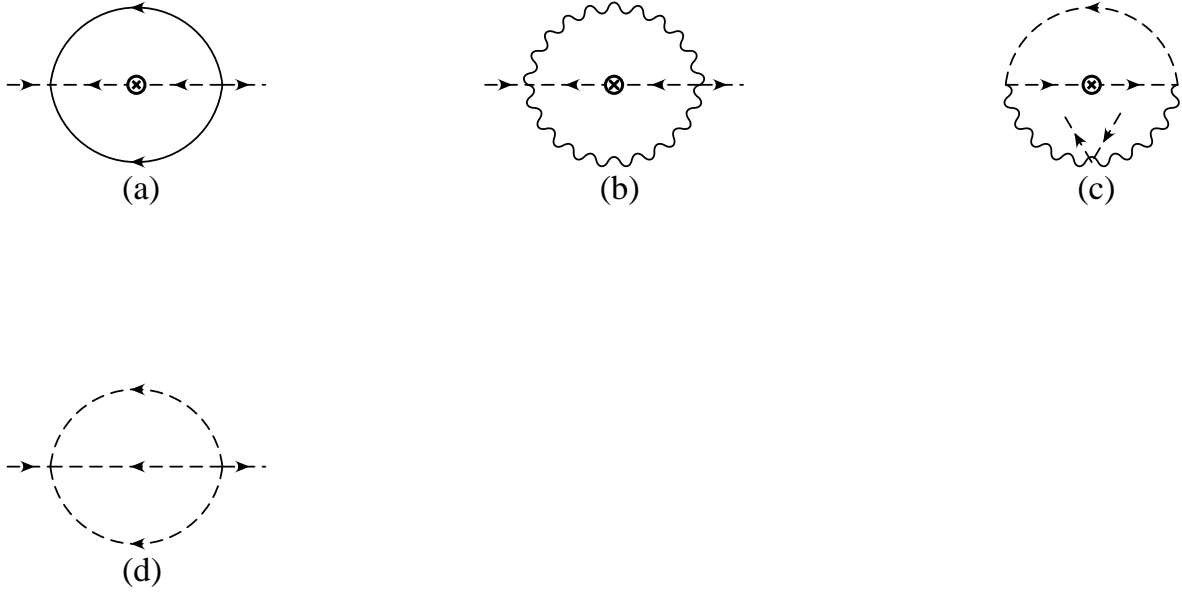}
\caption{Diagrams contributing to $\beta_{m^2}^{(2)}$}
\end{figure}

\begin{figure}
\includegraphics{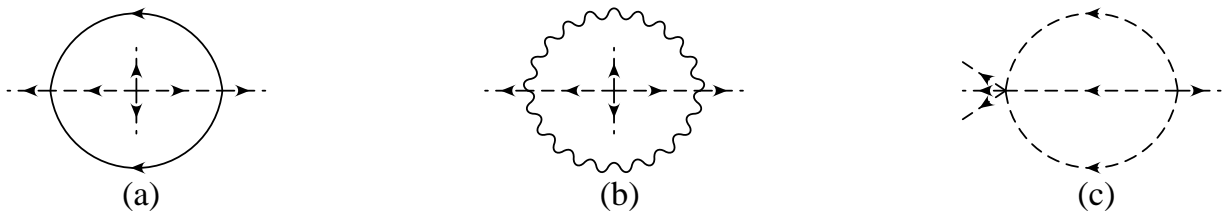}
\caption{Diagrams contributing to $\beta_h^{(2)}$}
\end{figure}

\end{document}